\documentclass[11pt]{article}
\usepackage[hmargin=2.5cm,vmargin=2cm]{geometry}
\usepackage{graphicx}
\usepackage{amsmath}
\usepackage{enumerate}

\begin{document}
\title{The thermodynamics of   self-gravitating systems in equilibrium is holographic}
\author { Ntina Savvidou\footnote{ksavvidou@physics.upatras.gr} and Charis Anastopoulos\footnote{anastop@physics.upatras.gr}  \\
 {\small Department of Physics, University of Patras, 26500 Greece} }
\maketitle

\begin{abstract}
We show that, when we study the coexistence of general relativity with thermodynamics, some physical properties that   are usually thought of as holographic and lying in the domain of quantum gravity can actually be accessed even at the classical level.
In particular,
we demonstrate that the thermodynamics of gravitating systems in equilibrium is  fully specified by variables defined on the system's boundary, namely, the boundary's geometry and extrinsic curvature.
Hence,  information is non-trivially incorporated in boundary variables because of the structure (the symmetries) of the classical gravity theory, without any input from quantum theory (such as black hole entropy).

\end{abstract}

\bigskip

\section{Introduction}
The association of entropy to black holes by Bekenstein and Hawking  strongly suggests of a fundamental relationship between gravity and thermodynamics. It has led to the conjecture that the fundamental theory of gravity should satisfy the holographic principle \cite{holog}, namely, the statement that the full information about a gravitational system is contained in the degrees of freedom of the system's boundary.

	The holographic principle refers to a mapping of bulk degrees of freedom into boundary degrees of freedom {\em at the Planck scale}.  In this work, we designate the equilibrium thermodynamics of self-gravitating systems as "holographic" because the thermodynamic state space consists solely  of {\em geometric} variables defined at the boundary; furthermore, measurements at the boundary suffice for establishing the thermodynamic properties of the system.  This is a restricted sense of holography, but it is highly non-trivial and it appears at the level of the coarse-grained quasiclassical description of an underlying quantum gravity theory---like, for example, in the proposals of Refs. \cite{Jac, Pad, Ver}. The most important feature of the holographic properties presented here is that they require
    no input from quantum theory   (such as black-hole entropy), let alone quantum gravity effects.  They originate from the structure of classical gravity theories and the basic principles of thermodynamics.

Our result follows from an important relation between Einstein's equations for static spacetimes and the principle of maximum entropy. Einstein's equations for a  general static spacetime  follows from the maximization of matter entropy subject to the continuity equation for the stress-energy tensor and  the initial value constraints---for an earlier proof of this statement, see  Ref. \cite{KM75};  Ref. \cite{old} for previous work and Refs. \cite{SWJ, Gao, new} for special cases. Here, we show that, for solutions to Einstein's equations, only boundary terms contribute to the total entropy of the system and that all thermodynamic properties are encoded into {\em geometric} variables on the boundary.

The structure of the article is as follows. In Sec. 2, we briefly describe the proof that the maximum entropy principle leads to Einstein's equations for static spacetimes and we compute explicitly the boundary term associated to variations of the appropriate thermodynamic potential. These results are employed in Sec. 3, in order to construct the thermodynamic state space of self-gravitating systems, explain the precise sense in which is holographic and set out the basic thermodynamic properties for self-gravitating systems. In Sec. 4, we  summarize our results and their implications.

\section{The maximum entropy principle in self-gravitating systems}
In this section, we show that the principle of maximum entropy for matter in the bulk implies that, for solutions to Einstein's equations, all thermodynamic variations can be expressed in terms of geometric properties of the boundary.
\subsection{Notation and definitions}
We  consider a static globally hyperbolic spacetime $M = R\times \Sigma$ with four-metric
\begin{eqnarray}
ds^2 = -L^2(x) dt^2 + h_{ij}(x) dx^i dx^j, \label{4metric}
\end{eqnarray}
expressed in terms of the spatial coordinates $x^i$ and the time coordinate $t$. The time-like vector field $\xi^{\mu} = (\partial/\partial t)^{\mu}$ is a Killing vector of the metric Eq. (\ref{4metric}). $L$ is the lapse function, and $h_{ij}$ is a $t$-independent Riemannian three-metric on the surfaces $\Sigma_t$ of constant $t$. The time-like   unit normal on $\Sigma_t$ is $n_{\mu}= L \partial_{\mu}t$ and the extrinsic curvature tensor on $\Sigma_t$ vanishes.

 Let  $C \subset \Sigma$ be a compact spatial region, with boundary $B = \partial C$. $C$ contains an isotropic fluid in thermal and dynamical equilibrium, described by the stress-energy tensor
\begin{eqnarray}
T_{\mu \nu} = \rho(x) n_{\mu}n_{\nu} + P(x) (g_{\mu \nu} + n_{\mu} n_{\nu}), \label{tumn}
\end{eqnarray}
where $\rho(x)$ and $P(x)$ are the energy density and the pressure, respectively.

 The stress-energy tensor satisfies the continuity equation $\nabla_{\mu}T^{\mu \nu} = 0$ which implies that
\begin{eqnarray}
\frac{\nabla_i P}{\rho +P} = - \frac{\nabla_iL}{L}. \label{cont}
\end{eqnarray}

 Eq. (\ref{tumn}) is the standard form of the stress-energy tensor for ideal fluids. Since dissipative processes are absent in equilibrium configurations,   Eq. (\ref{tumn}) also applies to non-ideal fluids.

We assume that the fluid consists of $q$ different particle species. The associated particle-number densities $n_a(x)$, $a = 1, \ldots, q$, together with the energy density $\rho(x)$ define the thermodynamic state space. We assume that all local thermodynamic properties of the fluid are encoded in the entropy-density functional $s(\rho, n_a)$. The first law of thermodynamics for the fluid takes the form
\begin{eqnarray}
T ds = d \rho - \sum_a \mu_a dn_a, \label{1st}
\end{eqnarray}
where $\mu_a = - T \frac{\partial s}{\partial n_a}$ is the chemical potential associated to particle species $a$ and $T = \left(\frac{\partial s}{\partial \rho}\right)^{-1}$ is the local temperature of the fluid. The pressure $P$ is defined through the Euler equation
\begin{eqnarray}
\rho + P - Ts -\sum_a \mu_a n_a = 0. \label{euler}
\end{eqnarray}
 Combining  Eqs. (\ref{euler}) and  (\ref{1st}), we derive the Gibbs-Duhem relation,
$dP  = s dT + \sum_a n_a d\mu_a$.

\subsection{ Entropy maximization}
Next, we maximize the total entropy of matter $S = \int_C d^3x \sqrt{h} s(\rho, n_a)$ for fixed values of the total particle numbers in $C$, $N_a = \int_C d^3x \sqrt{h} n_a(\rho, n_a)$ and for fixed values of the fields on the boundary.

Entropy maximization is subject to the continuity equation (\ref{cont}) for the fluid, and to the first-class  constraints of general relativity. For  static spacetimes, the Hamiltonian constraint reads
\begin{eqnarray}
{\cal H}(x) := 16\pi  \rho(x) - R(x) = 0, \label{Hamcon}
\end{eqnarray}
where $R$ is the Ricci scalar associated to the three-metric $h_{ij}$. The momentum constraint has been implemented in the choice of coordinates corresponding to the metric Eq. (\ref{4metric}), namely, the gauge-fixing condition that the shift vector vanishes.

 The Hamiltonian constraint expresses the energy density $\rho$ as a function of the metric $h_{ij}$ and the continuity equation expresses the lapse  $L$ in terms of the energy density $\rho$. It follows that the entropy $S$ and the particle numbers $N_a$ are functionals of the three-metric $h_{ij}$ and the particle-number densities $n_a$. Entropy maximization for fixed values of  $N_a$ requires the variation of $S + \sum_a b_a N_a$, with respect to $n_a$ and $h_{ij}$ for some Lagrange multipliers $b_a$.

Variation of  $S + \sum_a b_a N_a$ with respect to $n_a$ yields,
\begin{eqnarray}
 \delta S = \sum_a \int_C d^3x \sqrt{h} \left(\frac{\partial s}{\partial n_a} + b_a \right) \delta n_a = \sum_a \int_C d^3 x \sqrt{h} \left( -\frac{\mu_a}{T} + b_a\right) \delta n_a = 0,
\end{eqnarray}
leading to
 $b_a =  \frac{\mu_a}{T}$.  Hence, for equilibrium configurations the thermodynamic variables $\frac{\mu_a}{T}$ are constant in $V$.

We introduce the function $\omega$ (a Massieu function  \cite{Call}) as the Legendre transform of the entropy density $s$ with respect to $n_a$
\begin{eqnarray}
\omega(\rho, b_a) := s -\sum_a \frac{\partial s}{\partial n_a} n_a= s + \sum_a b_a n_a = \frac{\rho + P}{T}. \label{omega}
\end{eqnarray}

Expressed in terms of $\omega$, Eq. (\ref{1st}) takes the form
\begin{eqnarray}
 d \omega = \frac{d \rho}{T} - \sum_a n_a db_a.
\end{eqnarray}
It follows  that
$ T^{-1} = \partial \omega/\partial \rho$ and  $n_a = - \partial \omega/\partial b_a$. The Gibbs-Duhem relation becomes
$dP = \omega dT + T \sum_a n_a db_a$.

 Since the variables $b_a$ are constant for entropy-maximizing configurations,   $dP/dT = \omega = (P+\rho)/T$. Combining with Eq. (\ref{cont}), we obtain
\begin{eqnarray}
\frac{\nabla_iT}{T} = - \frac{\nabla_i L}{L}, \label{tol1}
\end{eqnarray}
which leads to Tolman's relation between local temperature and lapse function \cite{tolman}
\begin{eqnarray}
L T = T_*, \label{tolman}
\end{eqnarray}
where $T_*$ is  a constant. In an asymptotically flat spacetime, $L = 1 $ at spacelike infinity. Then, $T_*$ is identified with the temperature as seen by an observer at infinity.

\subsection{The thermodynamic boundary term}
Next, we show that for solutions to Einstein's equation, $\delta \Omega$ becomes a  boundary term.

 By employing the definition Eq. (\ref{omega}), entropy-maximization is equivalent to the maximization of the functional
\begin{eqnarray}
\Omega[h_{ij},b_a] := \int_C d^3x \sqrt{h} \omega(\rho,b_a). \label{OM}
\end{eqnarray}
Since the density $\rho$ depends on the metric $h_{ij}$ through the Hamiltonian constraint (\ref{Hamcon}) and $b_a$ are constant, $\Omega$ is a function on the space $Riem(C)$ of all Riemannian metrics $h_{ij}$ on $C$. Variation with respect to $h_{ij}$ yields

\begin{eqnarray}
\delta \Omega = \int_C d^3x \sqrt{h} \left( \frac{\omega}{2} h^{ij} \delta h_{ij} + \frac{\partial \omega}{\partial \rho} \delta \rho \right) = \int_C d^3x \sqrt{h} \left( \frac{\rho + P}{2T} h^{ij} \delta h_{ij} + \frac{\delta R}{16 \pi T} \right).
\end{eqnarray}
Using the equation $
\delta R = -R^{ij}\delta h_{ij} +\nabla^i \left(\nabla^j \delta h_{ij} - h^{kl} \nabla_i \delta h_{kl} \right)$ for the variation of the Ricci scalar $R$, together with Eq. (\ref{tol1}), we find
\begin{eqnarray}
\delta \Omega = - \int_C d^3 x \frac{\sqrt{h}}{16 \pi T} S^{ij} \delta h_{ij} + \frac{1}{16 \pi} \int_C d^3x \sqrt{h} \left[\nabla^i (T^{-1} \nabla^j \delta h_{ij} - T^{-1} h^{kl} \nabla_i \delta h_{kl}) \right. \nonumber \\
\left. - \nabla^j \left( \frac{\nabla^iL}{LT} \delta h_{ij} - \frac{\nabla_j L}{LT} h^{kl} \delta h_{kl}\right)\right],
 \label{domega}
\end{eqnarray}
where
\begin{eqnarray}
S^{ij} := R^{ij} - \frac{1}{2} h^{ij}R - \frac{1}{L} (\nabla^i \nabla^j L - h^{ij} \nabla_k \nabla^k L) - 8 \pi h^{ij}P.
\end{eqnarray}

The condition $S^{ij} = 0$ coincides the spatial components of Einstein's equations for the  static spacetime metric, Eq. (\ref{4metric}). For fixed values of the  metric $h_{ij}$ and of its first derivatives on the boundary $B$ of $C$, $\delta \Omega = - \int_C d^3 x \sqrt{h} S^{ij} \delta h_{ij} $.  Thus,
the principle of maximum entropy $\delta \Omega = 0 $ leads to full set of Einstein's equations \cite{KM75} for static spacetimes.

In order to properly interpret this result, we note that the equation $S^{ij} = 0$ corresponds to Hamilton's equations for the momentum $\pi^{ij}$, conjugate to $h_{ij}$ \cite{MTW}. In the Hamiltonian formulation of general relativity, the equations of motion incorporate the symplectic structure (Poisson bracket) and the constraints of the gravitational state space. The Hamiltonian constraint is employed in the derivation of Eq. (\ref{dom2}); and the momentum constraint is implemented in the specification of the metric Eq. (\ref{4metric}) and the gauge-choice of vanishing shift vector. It follows that the thermodynamic description of gravitating systems contains implicitly information about the symplectic structure of the theory \cite{Sav}.

However, the relation between the maximum entropy principle and Hamilton's equations, as incorporated in Eq. (\ref{domega}), is not specific to general relativity. It only requires the existence of first class constraints that relate the metric $h_{ij}$ and the energy density $\rho$ (for static configurations). Thus, it applies to a larger class of parameterized systems (i.e., systems whose Hamiltonian vanishes by virtue of first-class constraints). Consider, for example, a parameterized theory described   by the metric $h_{ij}$ and its conjugate momentum $\pi^{ij}$, and subject to a Hamiltonian constraint of the form
\begin{eqnarray}
{\cal H} =  T(\pi, h) + V(h) - \rho, \label{cons2}
\end{eqnarray}
 where $T$ is a `kinetic-energy' term and $V$ a `potential' term that may involve arbitrarily high derivatives of the three metric $h_{ij}$. Assume  that $T$ and $\delta T/\delta \pi^{ij}$ vanish for $\pi^{ij} = 0$. For this system, Hamilton's  equations for static solutions also follow from the maximum entropy principle.

The   total divergence term in  Eq. (\ref{domega})  determines the   thermodynamic properties of solutions to Einstein's equations. We evaluate this term by performing a $2+1$ decomposition of the metric $h_{ij}$ at the boundary $B$.

Let $B$ be described locally by the condition $f(x) = 0$. The unit normal to $B$ is $m_i = \alpha \nabla_i f$, where $\alpha = 1/\sqrt{h^{ij} \nabla_if \nabla_jf}$. The induced metric on $B$ is
\begin{eqnarray}
  \sigma_{ij} = h_{ij} - m_i m_j.
\end{eqnarray}
The tensor $\sigma^i_j = \delta^i_j - m^i m_j$ projects onto the two-surface $B$. Then, we define the
 extrinsic curvature tensor of $B$ as
\begin{eqnarray}
\kappa_{ij} := \sigma_{i}^k \sigma_{j}^l \nabla_k m_l. \label{ec}
\end{eqnarray}
 The metric and extrinsic curvature satisfy $\sigma_{ij} m^j = \kappa_{ij}m^j = 0$. Hence,   they are tensors on the 2-surface $B$.

Then, for solutions to Einstein's equations, Eq. (\ref{domega}) yields
\begin{eqnarray}
\delta \Omega =   \frac{1}{16 \pi T_*} \oint_{B} d^2y \sqrt{\sigma} \left(m^i \nabla_iL \sigma^{kl} \delta \sigma_{kl} - 2 L \sigma^{ij} \delta \kappa_{ij} + L \kappa^{ij} \delta \sigma_{ij}  \right), \label{dom2}
\end{eqnarray}
where the coordinates on $B$ are denoted as $y$.

A remarkable property of $\delta \Omega$, Eq. (\ref{dom2}), is that it depends only on the variations $\delta \sigma_{ij}$ of the two-metric and $\delta \kappa_{ij}$ of the extrinsic curvature $\kappa_{ij}$ of the boundary. The variations of the remaining components of the three-metric $h_{ij}$ and of its first derivatives do not contribute to $\delta \Omega$. This is a specific property of general relativity: the boundary term (\ref{dom2}) depends crucially on the form of the Hamiltonian constraint. The independence of $\delta \Omega$ from variations other than $\delta \sigma_{ij}$ and $\delta \kappa_{ij}$ does not follow from a generic first-class  constraint, such as Eq. (\ref{cons2}).

\section{Thermodynamics of self-gravitating systems}

\subsection{The thermodynamic state space}
In what follows, we argue that Eq. (\ref{dom2}) leads to holographic thermodynamics for self-gravitating systems in equilibrium. We employ the term "holographic thermodynamics" in the sense that the state space consists only of boundary variables and that all thermal properties of the system can be determined from local measurements at the boundary.

 Thermodynamic systems are defined in terms of a physical boundary that distinguishes the system from the rest of the universe. Equilibrium configurations are characterized by a number of quantities that are conserved in absence of external intervention; these quantities include the   internal energy $U$ and the particle-species numbers $N_a$.  In a flat background spacetime, the geometry of the boundary fully determines the enclosed volume $V$. Ordinary thermodynamic systems are extensive, i.e., their thermodynamic variables scale linearly with the volume $V$, so properties of the boundary other than the enclosed volume do not contribute to thermodynamics. In such systems, the state space $\Gamma$ of a thermodynamic system consists of the variables $(V, U, N_a)$, and the thermodynamic description requires the definition of an entropy functional on $\Gamma$.

The physical boundary $B$ of a self-gravitating system in equilibrium may correspond to a bounding box (as in ordinary thermodynamics) or to a stellar surface.  $B$ separates between a region in which the metric $h_{ij}$ is given by a solution of Einstein's equation with matter (the interior) and a region where $h_{ij}$ is a solution of the vacuum Einstein equations (the exterior). For stellar-surface boundary conditions, the pressure $P$ vanishes at $B$. When $B$ corresponds to a bounding box, the description of the spacetime geometry requires the use of junction conditions on $B$. According to the thin-shell formalism \cite{Israel, MTW},
  the variables $\sigma_{ij}$ and $\kappa_{ij}$ are continuous across $B$, while derivatives of the three-metric $h_{ij}$ normal to $B$ exhibit discontinuities proportional to the stress-energy tensor of the box. It is, therefore, significant that the only variables whose variations contribute to   $\delta \Omega$ in Eq. (\ref{dom2}) are the ones that remain continuous across $B$. If the variation of other variables, discontinuous across $B$, contributed to $\delta \Omega$, the thermodynamics would
    be strongly dependent on the physical properties of the box, as encoded in its boundary stress-energy tensor.

 An equilibrium configuration corresponds to a specific solution $h_{ij}$ of  Einstein's equations in the interior of $B$. According to Eq. (\ref{dom2}), only variations of the two-metric $\sigma_{ij}$ and of
  the extrinsic curvature $\kappa_{ij}$ have thermodynamic significance, i.e., their variations contribute to $\delta \Omega$. By Eq. (\ref{ec}),
  the extrinsic curvature can be viewed as a tangent vector on the space of Riemannian metric on $Riem(B)$, hence, the
    tensors   $\sigma_{ij}$ and   $\kappa_{ij}$ span the tangent bundle $T Riem(B)$.

    Viewed as boundary data for Einstein's equations $S^{ij}=0$ in the interior, $\sigma_{ij}$ and   $\kappa_{ij}$   do not suffice for finding a unique solution. Hence, in principle, many different solutions to Einstein's equations correspond to the same values of the thermodynamic variables on the boundary. Moreover, the integration of Einstein's equation from the boundary inwards may lead to singularities or other forms of internal boundaries---we give such examples in Sec. 3.2. In such a case, the variation of $\delta \Omega$, Eq. (\ref{dom2}) has to incorporate a contribution from the internal boundary which is, in general, inaccessible to an external observer.


Here, we restrict to the case of regular three-metrics $h_{ij}$, i.e., to three-metrics that are everywhere locally Minkowskian  in the interior of the physical boundary $B$. In this case,   Eq. (\ref{dom2}) involves integration only over the physical boundary of the system and no contributions from internal boundaries. Hence, {\em all variations in  Eq. (\ref{dom2}) can be attributed to   interventions by an external observer}, as is necessary for a consistent thermodynamic description of the system.

The thermodynamic properties of the system enclosed by $B$ are fully specified by the function $\Omega$, which depends on the surface fields
 $\sigma_{ij}$ and $\kappa_{ij}$
and on the variables $b_a$. The latter are constant in the interior, and thus their value is also determined at the boundary. It follows that the thermodynamic state space  for a self-gravitating system in equilibrium $\Gamma = T Riem(B) \times R^q$ consists only of variables that are accessible at the system's physical boundary.

As in ordinary thermodynamics, we define the  thermodynamic conjugate variables to $\sigma_{ij}$ and $\kappa^{ij}$ as
\begin{eqnarray}
A^{ij} := \frac{\delta \Omega}{\delta \sigma_{ij}} = \frac{1}{16 \pi T_*} \sqrt{\sigma}[(m^i \nabla_i L) \sigma^{ij} + L \kappa^{ij}], \hspace{2cm}
B^{ij} := \frac{\delta \Omega}{\delta \kappa_{ij}} = - \frac{1}{8 \pi T_*} L \sqrt{\sigma} \sigma^{ij}. \label{conjugate}
 \end{eqnarray}
 The conjugate variables above are tensor fields defined on the physical boundary $B$.   The particle numbers $N_a = - \frac{\partial \Omega}{\partial b_a}$ are the other thermodynamical conjugates and these are bulk variables. However, the particle numbers are conserved in a closed system, and their value can be ascertained from information about the process through which the system was assembled. An external observer with no knowledge about the system other than the conserved numbers $N_a$, can fully reconstruct the equations of state (the functional relation between state-space variables and their conjugates) solely by local measurements at the boundary. Thus, the thermodynamics of self-gravitating systems exhibit  striking manifestation of the holographic principle;
all hydrodynamic bulk variables (density, pressure) are encoded into geometric properties of the boundary.

\paragraph{Boundary temperature and pressure.} In ordinary thermodynamic systems, the conjugate variables on the thermodynamic state space correspond to the temperature and the pressure. In order to interpret the conjugate variables $A^{ij}$ and $B^{ij}$, Eq. (\ref{conjugate}), in terms of temperature and pressure, we consider
the projection of the equations of motion along the normal $m^i$ to the boundary $B$, $S_{ij}m^i m^j = 0 $. We obtain
\begin{eqnarray}
m^i \nabla_i L = \frac{L}{\kappa} \left[ 8 \pi P + \frac{1}{L} \sigma^{ij} {}^2\nabla_i {}^2\nabla_jL + \frac{1}{2} (\kappa_{ij}\kappa^{ij} - \kappa^2 +{}^2R) \right]. \label{dN}
\end{eqnarray}
where ${}^2 R$ is the curvature scalar on $B$, $\kappa = \kappa^{ij}\sigma_{ij}$, and ${}^2\nabla_i$ the covariant derivative associated to $\sigma_{ij}$.

 A natural thermodynamic assumption is that the local temperature is constant at the boundary $B$, ${}^2\nabla_i T = 0$, so that there is no boundary heat flow. This implies that ${}^2\nabla_iL = 0$ in Eq. (\ref{dN}). Denoting the temperature at the boundary as $T_B$, Eq. (\ref{conjugate}) leads to the equations of state for boundary pressure and momentum in a self-gravitating system
 \begin{eqnarray}
 \frac{1}{T_B} &=& -\frac{8 \pi}{\sqrt{\sigma}} \frac{\delta \Omega}{\delta \kappa_{ij}}\sigma_{ij} \label{tb} \\
 \frac{P}{T_B} &=& \frac{\kappa}{\sqrt{\sigma}} \frac{\delta \Omega}{\delta \sigma_{ij}} \sigma_{ij}-\frac{1}{32\pi T_B}(\kappa_{ij}\kappa^{ij} + \kappa^2 +{}^2R) \label{ptb}
 \end{eqnarray}

\paragraph{Boundary diffeomorphisms.} From Eqs. (\ref{tb}-\ref{ptb}) we observe that there are fewer functionally independent conjugate field variables (only the pressure  $P$) than variables of $\Gamma$. This is problematic for defining different thermodynamic representations of the system (such as the enthalpy or the free energy representations) using the Legendre transform. There are not enough conjugate variables to capture the information contained in the variables of $\Gamma$.

 This issue arises because we have not taken into account the diffeomorphism symmetry of the boundary, namely the fact that the thermodynamic description ought to be independent of the choice of coordinate system on the boundary $B$. Let us assume for concreteness that the boundary is topologically a two sphere: $B = S^2$. For any two-metric $\sigma$, there exists a coordinate system $\hat{y}^m$ on $S^2$, s.t. $\sigma_{mn} = e^{\phi} \hat{\sigma}_{mn}$, where $\phi$ is a scalar field and $\hat{\sigma}_{mn}$
 the homogeneous sphere metric of unit scalar curvature. Then, Eq. (\ref{dom2}) becomes
 \begin{eqnarray}
 \delta \Omega =  \frac{1}{16 \pi T_*} \oint_{B} d^2\hat{y} \sqrt{\hat{\sigma}}e^{\phi} \left[ (2 m^i\nabla_i L - L \kappa)\delta \phi - 2 L \delta \kappa\right] \label{dome}
 \end{eqnarray}
 The independent variations in Eq. (\ref{dome}) correspond to the conformal factor $\phi$ of the two-metric and the trace of the extrinsic curvature $\kappa$. This implies that the reduced state space $\Gamma_{red}$ of true thermodynamical degrees of freedom consists of $(\phi(y), \kappa(y), b_a)$ and it can be expressed as $\Gamma_{red} = T(Riem(B)/Diff(B))\times R^q$. Let us denote by $\pi: \Gamma \rightarrow \Gamma_{red}$ the projection map $\pi(\sigma_{ij}, \kappa_{ij}, b_a) = (\phi, \kappa, b_a)$. Then the thermodynamic function $\Omega$ on $\Gamma$ should be of the form $\Omega = \pi^*\tilde{\Omega}$, where $\tilde{\Omega}$ is a function on $\Gamma_{red}$.

 Eqs. (\ref{tb}-\ref{ptb})  imply that there exist two functionally independent fields on $B$ among the conjugate variables $\delta \Omega/\delta \phi(\tilde{y})$ and $\delta \Omega/\delta \kappa(\tilde{y})$. These functionally independent fields may be chosen as
   the pressure $P$ and the scalar $\kappa_{ab} \kappa^{ab}$. Hence, there is the same number of independent thermodynamically conjugate fields with the number of fundamental thermodynamic field $\phi$ and $\kappa$ on $\Gamma_{red}$. There is no problem of principle in defining   alternative thermodynamic representations through the Legendre transform of the function $\tilde{\Omega}$  on  $\Gamma_{red}$.

\subsection{ Non-regular solutions}
Next, we elaborate on the physical interpretation of non-regular solutions to Einstein's equations, and conjecture that their thermodynamic properties could lead to a much stronger version of holographic thermodynamics than the one derived here.

Our analysis so far is restricted to thermodynamic systems that correspond to regular solutions of Einstein's equations in the interior region. As explained earlier, the thermodynamic variables on the boundary under-determine the solution to Einstein's equations. Hence, it is in principle possible, that   the same values of the thermodynamic variables correspond to different spacetime geometries, even when restricting to regular solutions. In such a case, the holographic description refers only to the hydrodynamic properties of the interior and not the full physical description.

The consideration of the special case of spherical symmetric solutions to Einstein's equations clarifies this point. The interior solution corresponds to a three-metric
\begin{eqnarray}
ds^2_3 = \frac{dr^2}{1 - \frac{2m(r)}{r}} + r^2 (d\theta^2 + \sin^2\theta d \phi^2), \label{spheri}
\end{eqnarray}
in terms of the standard coordinates $(r, \theta, \phi)$; $m(r)$ is the mass function related to the density $\rho$, as $dm/dr = 4 \pi r^2 \rho$. The boundary $B$ corresponds to constant `radius' $r = r_0$. For $r > r_0$, the system is described by the Schwarzschild metric with mass $M$. Then, the two-metric and extrinsic curvature on $B$ are
\begin{eqnarray}
d \sigma^2 = r^2_0 (d\theta^2 + \sin^2\theta d \phi^2) \hspace{2cm} \kappa_{ij} = \frac{\sqrt{1 - 2M/r_0}}{r_0} \sigma_{ij}, \label{2sy}
\end{eqnarray}
i.e., they only depend on the two parameters $M$ and $r_0$, subject to the condition $r_0 > 2M$.
It follows that the thermodynamic state space $\Gamma$ of a spherically symmetric system is $(q+2)$-dimensional.

The interior metric $h_{ij}$ is a solution to the Oppenheimer-Volkoff (OV) equation. A unique solution to the OV equation  requires the determination of $r_0$, $m(r_0) = M$ and of the pressure $P(r_0)$ at the boundary, together with the variables $b_a$. Thus, the space $\Gamma_0$ of all solutions is $(q+3)$-dimensional, and $\Gamma$ is a submanifold of $\Gamma_0$, of codimension one.

A metric Eq. (\ref{spheri}) is regular if $m(0) = 0$; $m(0)$ may take no positive values for solutions to the OV equation \cite{ST}. For $m(0) = - M_0$, ($M_0 > 0$), the metric near $r = 0$ is approximated by

\begin{eqnarray}
ds^2_3 = \frac{r dr^2}{2 M_0} + r^2 (d\theta^2 + \sin^2\theta d \phi^2), \label{sing}
\end{eqnarray}

The proper radius coordinate corresponding to the metric Eq. (\ref{sing}) is $x = \frac{2}{3} r^{3/2}/\sqrt{2M_0}$. A two-sphere of proper radius $x$ around $r = 0$ has area equal to $4 \pi (\frac{9}{2} M_0)^{2/3} x^{4/3}$. This implies that the spacetime is not locally Minkowskian around $r = 0$ (since in that case the area should be $4 \pi x^2$), and that the point $r = 0$ is a conical singularity. This singularity does not give rise to inextensible causal geodesics, and it corresponds to an interior boundary of spacetime. Hence, the space of regular solutions $\Gamma_{reg}$ is also a submanifold of  $\Gamma_0$ of codimension one.

However, $\Gamma_{reg}$ does not coincide with the thermodynamic state space $\Gamma$. For some values of $(M, r_0, b_a)$ there exist more than one elements of $\Gamma_{reg}$, and for some other there exist none. It would seem that the thermodynamic state space $\Gamma$ should be restricted to the values of $(M, r_0)$ that correspond to a unique element of $\Gamma_{reg}$; otherwise the pressure at the boundary cannot be uniquely determined from the knowledge of $(M, r_0, b_a)$ and the equations of state are ambiguous.

In Ref.   \cite{AnSav12}, we argued that the restrictions described above are problematic for the conceptual consistency of gravitational thermodynamics. We proposed a resolution to this problem: a unique solution of Einstein's equation will be assigned to each element of $\Gamma$ through the principle of maximum entropy. Of all solutions to Einstein's equations characterized by the same boundary conditions (corresponding to an element of $\Gamma$), the physical solutions will be selected by the requirement that it maximizes the entropy functional. The implementation of this idea requires the assignment of an entropy term to the internal boundaries of  non-regular solutions, which we interpreted as a form of gravitational entropy to be added to the entropy of matter. We applied this idea successfully to a model system of spherically symmetric self-gravitating radiation in a bounding box \cite{AnSav12}. We found that the maximum entropy principle always selects a (unique) regular solution as an equilibrium configuration, provided a regular solution exists for given values of $(M, r_0)$; if a regular solution does not exist, then a non-regular solution is chosen uniquely. The generalization of this result to a larger class of spacetimes, and eventually to any self-gravitating system in equilibrium, is currently under investigation. This would lead to a much stronger version of the holographic principle than the one presented  here, namely that the thermodynamic variables at the boundary fully specify the geometry of the interior.

\subsection{Weak gravity limit}

At the weak gravity limit (i.e., for low densities and small system size), we expect that the thermodynamics of self-gravitating system reduces to ordinary thermodynamics expressed in terms of extensive variables.
In extensive thermodynamics, all variables are homogeneously distributed inside a bounding box $B$. Hence, at the weak gravity limit limit,
  the matter density $\rho$ is approximately constant on the interior of the   box. By Eq. (\ref{Hamcon}), the Ricci scalar $R$  will also be constant. Hence, the geometry inside the box  corresponds to a region of a homogeneous  three-sphere of radius $\sqrt{6/R}$.

  The thermodynamic properties of an extensive system are not affected by the shape of the bounding box; they depend only on the enclosed volume. Therefore, it suffices to consider a spherical boundary box $r = r_0$. It is convenient to  employ the spherical coordinates $(r, \theta, \phi)$ for the interior metric.
\begin{eqnarray}
ds^2 = \frac{dr^2}{1 - \frac{R r^2}{6}} + r^2 (d \theta^2 +\sin^2\theta d \phi^2). \label{3sph}
\end{eqnarray}
At the weak gravity limit, $R$ is close to zero, so Eq. (\ref{3sph}) is close the Minkowskian metric.

For a spherical bounding box $B$, the two-metric $\sigma_{ij}$ and the extrinsic curvature $\kappa_{ij}$ are given by Eq. (\ref{2sy}). Denoting by $\kappa_{ij}^{(0)}$ the extrinsic curvature of the surface $r = r_0$ when embedded in Minkowski spacetime, we find
\begin{eqnarray}
\Delta \kappa := \kappa - \kappa^{(0)} \simeq - \frac{R r_0^2}{12} \kappa^{(0)}.
\end{eqnarray}
to leading order in $R$, with $\kappa^{(0)} = - 2 /r_0$.
  Thus, the energy density $\rho$ of ordinary, extensive thermodynamics is proportional to the deviation of the extrinsic curvature from its Minkowskian value, while the box's geometry is kept constant. To leading approximation, the volume  enclosed by the box is $V_0 := \frac{4\pi }{3} r_0^3$ and the internal energy $U = \rho V_0$ of the system is
    \begin{eqnarray}
 U = - \frac{r_0}{12} \frac{\Delta \kappa}{\kappa^{(0)}}.
    \end{eqnarray}
 Hence, the function $\Omega(U, V_0, b_a) \simeq \omega( \rho, b_a) V_0$ is expressed solely in terms of the geometric variables $r_0$ and $\Delta \kappa$ of the boundary. Hence, the theory of ordinary extensive thermodynamics can be reformulated in terms of the Riemannian geometry of the three-sphere.

 Finally, we note that an alternative characterization for $U$ can be made in terms of the actual volume $V$ enclosed by the spherical boundary, whence
 \begin{eqnarray}
 U = \frac{3 r_0}{5} \frac{\Delta V}{V_0},
 \end{eqnarray}
 where $\Delta V = V - V_0$.

 We must emphasize here that, the spacetime curvature persists even at the weak gravity limit and it is essential for the thermodynamic description in terms of boundary variables. The holographic properties presented here are not accessible in the Newtonian theory of gravity.

\subsection{Thermodynamic properties}

Next, we formulate the basic laws of thermodynamics in light of the previous results, we examine possible candidates for the notion of internal energy in self-gravitating systems and we derive non-trivial thermodynamic inequalities.

\paragraph{Zero-th law of thermodynamics.}  The generalization of the zero-th law of thermodynamics to self-gravitating systems is provided by Tolman's relation between lapse function and local temperature,  Eq. (\ref{tolman}). Indeed, for an asymptotically flat spacetime,  Eq. (\ref{tolman}) is equivalent to the statement that the temperature of a self-gravitating body in equilibrium is everywhere constant when transformed to the frame of an observer   at infinity.

 From the analysis of Sec. 2, we note that the only assumptions required for the derivation of  Eq. (\ref{tolman}) are (i) that the total entropy is expressed in terms of an entropy  density $s$, and (ii) that the Hamiltonian constraint does not involve the particle densities $n_a$.  Tolman's relation is essentially kinematical, in the sense that  its derivation does not depend on the specific form of Einstein's equations---it holds for any theory described by a constraint of the form Eq. (\ref{cons2}). In effect, Tolman's law follows from Galileo's principle  that gravity   is insensitive to the physical composition of material bodies.

 We also note the inter-relation between the zero-th law of thermodynamics and   the continuity equation, Eq. (\ref{cont}). Either of the two can be assumed as {\em a priori} given, and the other will be derived   as a consequence of the maximum entropy principle.

\paragraph{ First law of thermodynamics.} Including the contribution from variations to $b_a$ in Eq. (\ref{dom2}) and using
    Eq. (\ref{dN}), we derive   is the first law of thermodynamics for a self-gravitating system,
\begin{eqnarray}
 \delta \Omega &=&  \frac{1}{T_*} \oint_B d^2y \frac{LP}{\kappa}\delta(\sqrt{\sigma}) + \sum_a N_a \delta b_a \nonumber \\
&+& \frac{1}{16 \pi T_*} \oint_B d^2y L \sqrt{\sigma} \left(\kappa^{ij}\delta \sigma_{ij} - 2 \sigma^{ij} \delta \kappa_{ij} + \frac{\kappa_{ij}\kappa^{ij} - \kappa^2 + {}^2R}{2 \kappa} \sigma^{kl} \delta \sigma_{kl}\right) , \label{1law}
\end{eqnarray}
or, using the entropy functional $S$ instead of $\Omega$
\begin{eqnarray}
 \Theta = T_* dS  - \oint_B d^2y\frac{LP}{\kappa}\delta(\sqrt{\sigma}) + \sum_a (L \mu_a)_B \delta N_a, \label{1law2}
\end{eqnarray}
where we denoted   the second line of Eq. (\ref{1law}) by $\Theta/T_*$; $ \Theta$ is an one-form on the thermodynamic state space $\Gamma$. The terms in the right-hand-side of Eq. (\ref{1law2}) have an obvious thermodynamic significance, as heat, mechanical work and `chemical' work. However, $\Theta$ is not an exact form: $\delta \Theta \neq 0$. So, $\Theta$ cannot be written as $\delta U$, where $U$ is a scalar function on $\Gamma$, the system's internal energy. $\Theta$ also contains work terms.

There is no obvious definition of internal energy for a general self-gravitating system. This is due to the fact that the maximum entropy principle leading to Eq. (\ref{dom2}) does not require the specification of constant internal energy \cite{KM75}. Candidates for the internal energy such as the Komar mass (see below) are boundary terms; hence, they do not affect the derivation of Einstein's equations from the maximum entropy principle. The internal energy   is not an independent variable in the thermodynamic state space of self-gravitating systems---unlike ordinary thermodynamics---but a function of the more fundamental geometric variables $\sigma_{ij}$ and $\kappa_{ij}$.

For a spherically symmetric system, the variables $\sigma_{ij}$ and $\kappa_{ij}$ are given by Eq. (\ref{2sy}), and  $L =\sqrt{1 - 2M/r_0}$. The one-form $\Theta$ becomes $\Theta =  \delta M$, and the first law of thermodynamics simplifies

\begin{eqnarray}
\delta M = T_* \delta S - P(r_0) (4 \pi r_0^2 \delta r_0) + \sqrt{1 - 2M/r_0} \sum_a \mu_a(r_0) \delta N_a. \label{sph2}
\end{eqnarray}

 Hence, in spherically symmetric systems,  internal energy is identified with the Arnowitt-Deser-Misner mass $M$, and the one-form $\Theta$ contains no work terms. This suggests that, in the general case, the work terms in $\Theta$ are associated to inhomogeneities of the boundary.

\paragraph{The Komar mass.}
The leading candidate    for the internal energy is the Komar mass $M$,  defined in terms of the surface integral \cite{gour}
\begin{eqnarray}
M = \frac{1}{4 \pi} \oint_{B'} d^2 y\sqrt{\sigma}  m^k \nabla_k L. \label{Komar}
\end{eqnarray}
for any closed surface $B'$ enclosing $B$, since the energy density and pressure vanishes in the exterior of $B$. The Komar mass is the conserved quantity associated to the symmetry of time translations in a static spacetime via Noether's theorem, so it is natural to identify it with the internal energy. However, there are two problems to such an identification.  First, there is no obvious physical interpretation of  $\Theta - \delta M$  as a work term. Second, for a system in a bounding box, the variable $m^k \nabla_k L$ appearing in Eq. (\ref{Komar}) is discontinuous across the boundary. Hence, the Komar mass may turn out to depend on the stress-energy tensor of the bounding box, and not only on the properties of the enclosed fluid. This is unacceptable for the internal energy of a thermodynamic system.

Irrespective of its status as a candidate for internal energy, the Komar mass is closely related to the thermodynamic function $\Omega$. For a gravitating system bound  by a box, the pressure is discontinuous at $B$. It is non-zero inside the box, but vanishes outside. The metric components and the derivatives corresponding to the thermodynamical variables  $\sigma_{ij}$ and $\kappa_{ij}$ are continuous across the boundary, but the derivative $m^k \nabla_k L$ exhibits a jump due to the discontinuity of the pressure. We denote the value of    $m^k \nabla_k L$ in the outside of $B$ with the suffix $+$ and its value inside $B$ by the suffix $-$.   From Eq. (\ref{dN}), we obtain
\begin{eqnarray}
(m^k \nabla_k L)_+ - (m^k \nabla_k L)_- = - 8\pi \frac{L P}{\kappa}.
\end{eqnarray}

 Taking the limit of $B' \rightarrow B$ from the outside,  Eq. (\ref{Komar}) becomes
\begin{eqnarray}
M =  - 2  \oint_B d^2 y\sqrt{\sigma} \frac{L P}{\kappa} + T_* \int_C d^3 x  \sqrt{h}  \frac{\rho+ 3 P}{T}, \label{km0}
\end{eqnarray}
In deriving Eq. (\ref{km0}), we employed the equation
\begin{eqnarray}
4 \pi (\rho + 3 P) L = \nabla_k \nabla^k L,
\end{eqnarray}
 that follows from $S_{ij}h^{ij} = 0$, and Tolman's law, Eq. (\ref{tolman}). Using Eq. (\ref{OM}), we find
\begin{eqnarray}
2 \int_C d^3 x  \sqrt{h}  \frac{\rho}{T} = 3 \Omega - \frac{ M + 2 \oint_B d^2 y\sqrt{\sigma} L P/\kappa}{T_*}. \label{key}
\end{eqnarray}

 Eq. (\ref{key}) is the a result relating the Komar mass $M$ and the thermodynamic function of $\Omega$, because the bulk term in the l.h.s. of Eq. (\ref{key}) can often be explicitly related to $\Omega$. Consider, for example, a fluid with a linear equation of state $P = \gamma \rho$, where $\gamma$ is a constant. For such a fluid
  \begin{eqnarray}
  \Omega = (1+\gamma) \int_C d^3 x  \sqrt{h}  \rho/T ,
\end{eqnarray}

   and, Eq. (\ref{key}) becomes
\begin{eqnarray}
\Omega = \frac{1 + \gamma}{1 + 3 \gamma}\frac{ M + 2 \oint_B d^2 y\sqrt{\sigma} L P/\kappa}{T_*}. \label{lin}
\end{eqnarray}
 Eq. (\ref{lin}) demonstrates explicitly how $\Omega$ is expressed solely in terms of   variables that are defined on the boundary $B$. For  spherically symmetric systems, Eq. (\ref{lin}) reproduces the results of Ref. \cite{SWJ}.

\paragraph{Thermodynamics inequalities.} An important motivation for formulating gravitational thermodynamic in terms of boundary variables is the possibility of formulating the entropy bounds suggested by black hole thermodynamics in an invariant geometric language \cite{enb} which can also be used for studying their implications to ordinary. While a proof of such entropy bounds requires a significant extension of the present formalism, Eq. (\ref{key}) does lead to several non-trivial thermodynamic inequalities.

  For a fluid that satisfies the weak and dominant energy conditions ($0 \leq P \leq \rho$),
  \begin{eqnarray}
  \frac{1}{2}\Omega  \leq \int_C d^3 x  \sqrt{h}  \rho/T \leq \Omega.
  \end{eqnarray}

   Hence, by Eq. (\ref{key})
\begin{eqnarray}
\frac{ 1}{2 } \leq \frac{\Omega T_*}{M + 2 \oint_B d^2 y\sqrt{\sigma} L P/\kappa}  \leq 1. \label{ineq}
\end{eqnarray}

For stellar-surface boundary conditions, $P = 0$,  and Eq. (\ref{ineq}) simplifies
\begin{eqnarray}
 \frac{M}{2T_*} \leq \Omega \leq \frac{M}{T_*}. \label{bound}
\end{eqnarray}

Interestingly,  the lower bound to $\Omega$ in Eq. (\ref{bound}) is saturated by a black hole. Strictly speaking the present formalism does not apply to black holes. However, in a heuristic sense, we can model
  a Schwarzshild black hole as a thermodynamical system with vanishing chemical potential (because the particle numbers are not preserved due to the no-hair theorem) that is characterized by the Bekenstein-Hawking entropy $S = 4 \pi M^2/\hbar$ and by the Hawking temperature $T_* = \hbar/(8 \pi M)$. Then,  we readily verify that $S = \Omega =  \frac{M}{2T_*}$.

A different bound to $\Omega$ follows by noting that in a stellar solution, the temperature $T$ in the interior is larger than the surface temperature and consequently $T \leq T_*$ (this property does not hold for non-regular solutions). Eq. (\ref{key}) then becomes
\begin{eqnarray}
2 M_P \leq 3 \Omega T_* - M, \label{mpin}
\end{eqnarray}
 where $M_P = \int_C d^3 x  \sqrt{h} \rho$ is the {\em proper mass} of the fluid in the interior. The difference between $M$ and $M_P$ is interpreted as the gravitational binding energy $E_B$ of the configuration
 \begin{eqnarray}
 E_B = M_P - M.
  \end{eqnarray}
Then, Eq. (\ref{mpin}) becomes
\begin{eqnarray}
 M - \Omega T_* \leq - \frac{2}{3} E_B,
\end{eqnarray}
i.e., the binding energy defines an upper bound to the generalized free energy $M - \Omega T_*$.

\section{Conclusions.}

The main result of this article is the demonstration that the thermodynamics of gravitating systems in equilibrium is holographic at the level of classical general relativity, in a very precise sense: (i) the thermodynamic state space consists of variables defined on the boundary, and (ii) the thermodynamic properties of the system are ascertained by local measurements at the boundary. We believe that these results can be extended towards a stronger sense of holography. To this end, we showed in Ref. \cite{AnSav12} that the full geometry in the bulk follows from the knowledge of thermodynamic variables at the boundary in a specific self-gravitating system, and we believe that this result can be generalized.

We showed that the definition of thermodynamics in terms of boundary variables is a property of  parameterized theories, in general, not of general relativity in particular. However, general relativity is distinguished because it leads to a geometrically natural state space, that consists of the boundary two-metric and its extrinsic curvature. Thus, the equilibrium thermodynamics of gravitating systems can be described in a geometrical language, and this property persists at the  limit of ordinary extensive  thermodynamics.

We view these results as an important stepping stone towards the formulation of a general axiomatic theory of equilibrium thermodynamics in gravitational systems that will also include black holes. At the moment, the major open issues are: (i) the thermodynamic properties of the interior boundaries that characterize non-regular solutions, and (ii) a thermodynamically natural definition of internal energy.

\end{document}